\documentclass[aps,groupedaddress,superscriptaddress,amsmath,amssymb,pra,twocolumn]{revtex4-2}
\usepackage{lipsum}
\usepackage{dcolumn}
\usepackage{bm}
\usepackage[hidelinks]{hyperref}
\usepackage{float}
\usepackage{graphicx}
\usepackage{amsfonts}
\usepackage{verbatim}
\usepackage{bbm}
\usepackage{braket}
\usepackage{qcircuit}
\usepackage[normalem]{ulem}
\usepackage[T1]{fontenc} 
\hypersetup{
    colorlinks=true,
    linkcolor=blue,
    filecolor=magenta,      
    urlcolor=cyan,
}

\usepackage{color}
\usepackage{multirow}
\usepackage{array}
\usepackage{tabularx}
\usepackage{cancel}
\def\ep{\varepsilon}

\def\xb{{\bf x}}
\def\yb{{\bf y}}

\def\h1{\mathds{1}}

\draft

\begin{document}
%\linenumbers

\title{Frequency stabilization of self-sustained oscillations in a sideband-driven electromechanical resonator}
\author{B. Zhang}
\affiliation{Department of Physics, the Hong Kong University of Science and Technology, Clear Water Bay, Kowloon, Hong Kong, China}
\affiliation{William Mong Institute of Nano Science and Technology, the Hong Kong University of Science and Technology, Clear Water Bay, Kowloon, Hong Kong, China}
\author{Yingming Yan}
\affiliation{Department of Physics, the Hong Kong University of Science and Technology, Clear Water Bay, Kowloon, Hong Kong, China}
\affiliation{William Mong Institute of Nano Science and Technology, the Hong Kong University of Science and Technology, Clear Water Bay, Kowloon, Hong Kong, China}
\author{X. Dong}
\affiliation{Department of Physics, the Hong Kong University of Science and Technology, Clear Water Bay, Kowloon, Hong Kong, China}
\affiliation{William Mong Institute of Nano Science and Technology, the Hong Kong University of Science and Technology, Clear Water Bay, Kowloon, Hong Kong, China}
\author{M. I. Dykman}
\email{dykmanm@msu.edu}
\affiliation{Department of Physics and Astronomy, Michigan State University, East Lansing, Michigan 48824, USA}
\author{H. B. Chan}
\email{hochan@ust.hk}
\affiliation{Department of Physics, the Hong Kong University of Science and Technology, Clear Water Bay, Kowloon, Hong Kong, China}
\affiliation{William Mong Institute of Nano Science and Technology, the Hong Kong University of Science and Technology, Clear Water Bay, Kowloon, Hong Kong, China}

\date{\today}

\begin{abstract} 

We present a method to stabilize the frequency of self-sustained vibrations in micro- and nanomechanical resonators. The method refers to a two-mode system with the vibrations at significantly different frequencies. The signal from one mode is used to control the other mode. In the experiment, self-sustained oscillations of micromechanical modes are excited by pumping at the blue-detuned sideband of the higher-frequency mode. Phase fluctuations of the two modes show near perfect anti-correlation. They can be compensated in either one of the modes by a stepwise change of the pump phase. The phase change of the controlled mode is proportional to the pump phase change, with the proportionality constant independent of the pump amplitude and frequency. This finding allows us to stabilize the phase of one mode against phase diffusion using the measured phase of the other mode. We demonstrate that phase fluctuations of either the high or low frequency mode can be significantly reduced. The results open new opportunities in generating stable vibrations in a broad frequency range via parametric downconversion in nonlinear resonators.

\begin{description}
\item[Subject Areas]
Mechanics, Nonlinear Dynamics, Nanophysics
\end{description}

\end{abstract}

\maketitle

\section{Introduction}
\label{sec:one}

Self-sustained oscillations play  a major role in diverse fields ranging from biological systems to lasers and clocks \cite{jenkins_self-oscillation_2013}. One of the most important classes of systems that can display self-sustained oscillations is mechanical resonators. Examples include quartz oscillators that have long served as time references in ordinary clocks. In recent years, silicon micromechanical devices \cite{kim_frequency_2007,van_beek_review_2012,zaliasl_3_2015,roshan_111_2016-1}, because of their small size, have become viable alternatives of quartz oscillators in numerous applications, from telecommunications to smart watches. Self-oscillating micro- and nanomechanical devices are also used in high-precision measurements. Shifts in the vibration frequency enable measuring force \cite{eichler_symmetry_2013}, charge \cite{cleland_nanometre-scale_1998}, spin \cite{rugar_single_2004} and mass \cite{hanay_single-protein_2012}. A well-known example is frequency modulation atomic force microscopy \cite{Albrecht1991}. The decrease of the size of mechanical resonators is generally expected to lead to an increase of fluctuations \cite{Vig1999}. Suppressing fluctuations is a central problem in micro- and nanomechanical systems, and in particular in the systems that display self-sustained oscillations.

In micromechanical devices, the feedback needed to excite self-sustained oscillations is often implemented via an external circuit in which the signal generated by vibrations is phase-shifted and amplified to serve as the periodic drive \cite{PhysRevA.51.4211,feng_self-sustaining_2008,kenig_optimal_2012,villanueva_nanoscale_2011,antonio_frequency_2012,chen_self-sustained_2016,huang_frequency_2019-1,Defoort2021}. Alternatively, the feedback mechanism can also be intrinsically built into the driven system, such as by photothermal  \cite{velikovich_tristability_1991,zalalutdinov_autoparametric_2001,Barton2012,de_alba_low-power_2017,roxworthy_electrically_2018} or other effects \cite{ochs_frequency_2022-1},  or by measurement backaction \cite{etaki_self-sustained_2013}. 

Phase fluctuations in self-sustained oscillations are induced by different sources, from the thermal noise that arise from the coupling of the resonator to the environment described by the fluctuation-dissipation theorem to noise in various parts of the feedback circuit \cite{Schmid2016,Demir2020,Bachtold2022a}. There have been many efforts to improve the phase stability of self-sustained oscillations in micro- and nano-mechanical systems. In the linear regime, phase fluctuations induced by thermal noise decreases with the oscillation amplitude \cite{Rytov1956,Rytov1956a,Rubiola2009}. It is therefore beneficial to maximize the oscillation amplitude by operating the feedback circuit with large external drive. However, if the oscillation amplitude is increased beyond the linear regime, the eigenfrequency of the resonator becomes dependent on the amplitude. As a result, amplitude fluctuations are converted into frequency fluctuations. It was shown that, by designing the dependence of the eigenfrequency on amplitude to be non-monotonic, phase noise can be minimized at the extremum where the eigenfrequency is locally independent of amplitude \cite{Huang2019,Miller2021}.

The nonlinear coupling to high order modes has also been exploited to reduce phase noise. In these systems, the two modes are tuned into internal resonance, with the ratio of the eigenfreqeuncies of the modes being an integer. With energy exchanged between the modes, the phase fluctuations of the self-sustained oscillations of the lower mode become much weaker \cite{Antonio2012,Zhao2017}.

Other approaches focus on mitigating the effects of noise generated by the feedback loop. For example, noise in the phase of the periodic drive can produce phase noise in the oscillations. This effect can be significantly reduced by driving a resonator into nonlinearity and picking an operation point where the vibration frequency is insensitive to the phase of the drive \cite{Greywall1994,Yurke1995}. More recent works extended these concepts \cite{Kenig2012a,Kenig2013} and explored the possibility of surpassing the limit imposed by the thermomechanical noise \cite{Villanueva2013}. Similar approaches were also used in minimizing the phase noise in self-sustained oscillations of two vibrational modes under nondegenerate parametric drive \cite{Kenig2012}. 

An important distinction of self-sustained oscillations from forced oscillations is that the oscillation phase is arbitrary. In the presence of noise, fluctuations of the phase accumulate in time, resulting in phase diffusion \cite{Berstein1938,Rytov1956,Haken1966,Lax1967,Agrawal2014,sun_correlated_2016,Demir2000}. Phase diffusion leads to decoherence, determining the spectral width of self-sustained oscillations. Minimizing phase diffusion is therefore paramount in applications of self-sustained oscillations ranging from high sensitivity detection to high stability frequency standards. To our knowledge, the possibility of significant reduction of phase diffusion in self-sustained oscillations has neither been explored theoretically nor demonstrated experimentally in micro- and nano-mechanical oscillators.

In this paper, we present a scheme to stabilize the frequency of an important type of self-sustained oscillations induced by a drive with a frequency that is orders of magnitude different from the frequency of the vibrations of interest. We consider the vibrations induced by dynamical backaction in a sideband-driven micromechanical resonator with two nonlinearly coupled vibrational modes. The two modes have vastly different eigenfrequencies and decay rates, with the higher mode serving as an analog of a photon cavity mode \cite{sun_correlated_2016,mahboob_phonon-cavity_2012} in the context of cavity optomechanics \cite{kippenberg_cavity_2008,aspelmeyer_cavity_2014}. Sufficiently strong pump applied at the blue-detuned sideband of the upper mode excites self-sustained vibrations \cite{dykman1978heating,marquardt_dynamical_2006,kippenberg_analysis_2005,carmon_temporal_2005,metzger_self-induced_2008,bagheri_dynamic_2011,faust_microwave_2012,buters_experimental_2015} of both modes. In the stationary regime, the phase diffusion of the modes is anti-correlated. The sum of the phases remains essentially constant \cite{sun_correlated_2016}. 

We show that the sum of the phases can be adjusted by the phase of the sideband pump and study the transient process in which such adjustment occurs when the pump phase is changed. For a step change of the pump phase, the phases of both modes settle to new values after a transient, the duration of which is determined by the smallest eigenvalue of the linearized equations of motion about the stable state. The phase change of each mode is proportional to the pump phase change, which is assumed to be small. Importantly, the two proportionality constants add up to unity. This finding, together with the phase anti-correlation of the two modes, allow us to stabilize the phase of one mode by measuring the phase of the other mode and then compensating for the phase diffusion by adjusting the phase of the pump. We demonstrate that phase fluctuations of either the high or low frequency mode can be significantly reduced. This results in a much narrower spectral linewidth. Our scheme does not require a frequency reference near the mode to be stabilized, in contrast to direct feedback that stabilizes a particular mode by measuring its phase. By exploiting the nonlinear coupling between two modes, the frequency reference could be orders of magnitude different from the eigenfrequency of the mode that is stabilized.

\begin{figure}
\includegraphics[scale=0.5]{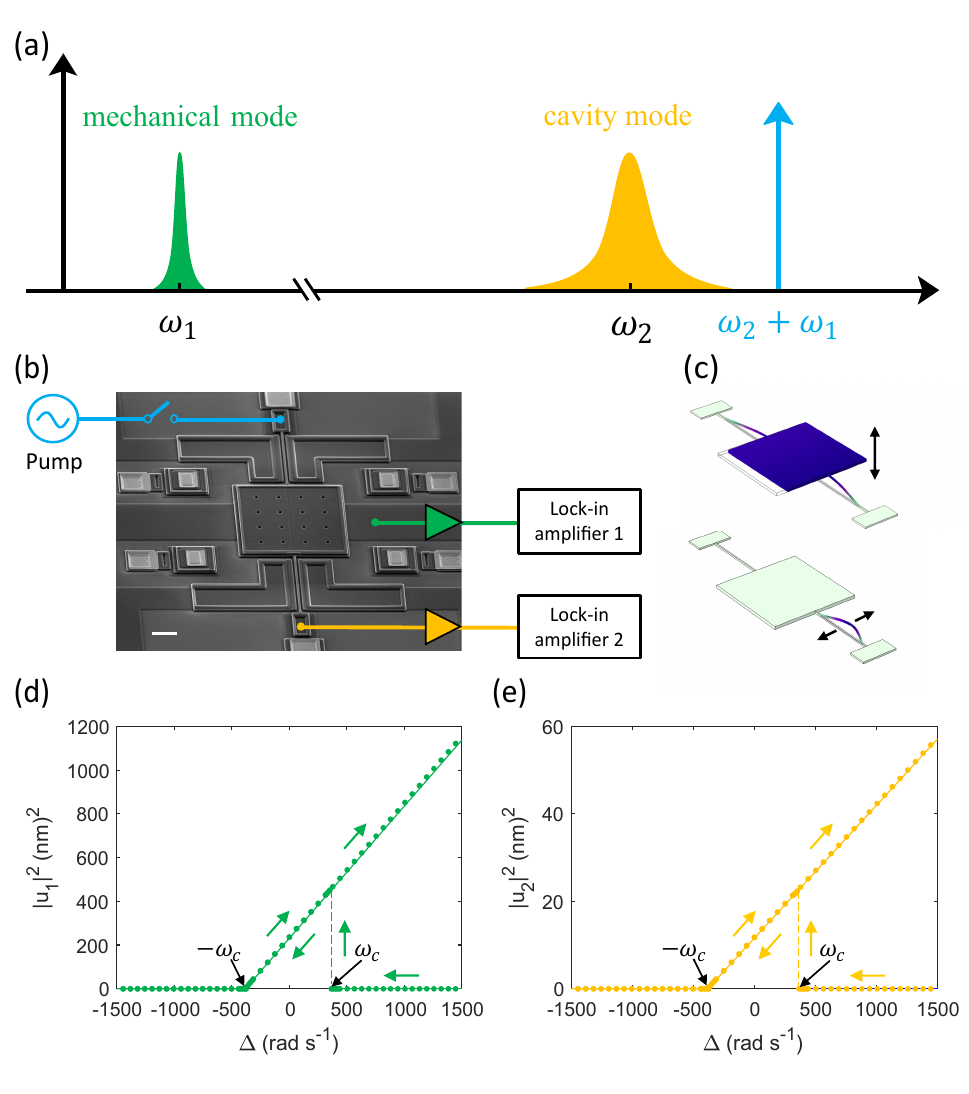}

\caption{(a) A sketch of the spectra of the mode 1 (mechanical mode) and mode 2 (cavity mode). Energy of the pump that leads to self-sustained vibrations is supplied to the upper sideband of mode 2. (b) Scanning electron micrograph of electromechanical resonator including the measurement scheme. The white scale bar at the lower left corner measures 30 $\mu$m. (c) Vibration profiles of mode 1 (top) and mode 2 (bottom). (d) Dependence of the square of vibration amplitude of mode 1 on the detuning of the pump frequency from the upper sideband for mode 1 at pump current amplitude of 189 $\mu$A. (e) Similar plot for mode 2.}

\label{Fig.1} 
\end{figure}

\section{TWO-MODE ELECTROMECHANICAL SYSTEM}
\label{sec:two}

\subsection{Excitation and detection of self-sustained vibrations}
\label{subsec:2A}

In our experiment, the micromechanical resonator consists of a square polysilicon plate ($100\,\mu$m$\,\times\,100\,\mu$m$\,\times\,3.5\,\mu$m) that is suspended by two beams of the same cross-section area ($1.3\,\mu$m$\ \mathrm{by} \ 2\,\mu$m) but different lengths of $80\,\mu$m and $75\,\mu$m respectively [Fig.~\ref{Fig.1}(b)]. The beams are covered by 30 nm of gold to reduce the electrical resistance. We utilize two vibrational modes in our experiment. For the low frequency mode (mode 1), the plate undergoes translational motion normal to the substrate as the two beams are deformed [top of Fig.~\ref{Fig.1}(c)]. The resonance frequency $\omega_1/2\pi$ is 47030.7 Hz and the damping constant $\Gamma_1/2\pi$ is 0.48 Hz. The high frequency mode (mode 2) is the lowest in-plane mode of the shorter beam [bottom of Fig.~\ref{Fig.1}(c)], with resonant frequency $\omega_2/2\pi=1867195.4$ Hz and damping constant $\Gamma_2/2\pi=36.2$ Hz. Because of its significantly larger decay rate, in the presence of a weak external drive mode 2 can play a role similar to optical/microwave modes in cavity optomechanics \cite{kippenberg_cavity_2008,aspelmeyer_cavity_2014}. The change of the tension caused by the vibrations leads to a nonlinear coupling between the modes. 

Forced vibrations of each mode can be excited by periodic electrostatic forces when small probe ac voltages are applied to the electrodes nearby. For mode 2, vibrations can also be excited by the periodic Lorentz force when an ac current is passed through the beams in a magnetic field perpendicular to the substrate. Motion of the plate in mode 1 is detected by measuring the capacitance change between the plate and the underlying electrode. Vibrations in mode 2 are detected by the change in the current due to the back electromotive force as the beam vibrates in the magnetic field. All measurements are carried out at a temperature of 4 K and pressure of $<10^{-5}$  torr. These measurements allowed us to determine the values of the eigenfrequencies and decay rates of the modes.

To induce self-sustained oscillations, a periodic pump current at frequency $\omega_F$ close to the frequency sum ($\omega_1+\omega_2$) is applied through the beam to generate a Lorentz force as shown in Fig.~\ref{Fig.1}(a). This force modulates the nonlinear coupling between the two modes. The equations of motion that describe the effect of such resonant modulation on the coupled-mode system are: 
\begin{align}
\label{eq1}
%&\MD{\ddot q_i+\omega_i^2q_i+2\mathrm{\Gamma}_i \dot q_i+\frac{\gamma_i}{m_i}q_i^3+\frac{\gamma}{m_i}%q_iq_{3-i}^2 }
%\nonumber\\
%& \MD{ =\frac{F}{m_i}q_{3-i}\cos\left(\omega_Ft+\theta_F\right), \quad (i=1,2) }\nonumber\\
&{\ddot{q}}_1+\omega_1^2q_1+2\mathrm{\Gamma}_1{\dot{q}}_1+\frac{\gamma_1}{m_1}q_1^3+\frac{\gamma}{m_1}q_1q_2^2 
\nonumber\\
&
 =\frac{F}{m_1}q_2cos{\left(\omega_Ft+\theta_F\right),} \nonumber\\
&{\ddot{q}}_2+\omega_2^2q_2+2\mathrm{\Gamma}_2{\dot{q}}_2+\frac{\gamma_2}{m_2}q_2^3+\frac{\gamma}{m_2}q_2q_1^2 \nonumber\\
& =\frac{F}{m_2}q_1cos{\left(\omega_Ft+\theta_F\right),}
\end{align}
where $q_{1,2}$ are the displacements of the plate and the beam center, respectively, $m_{1,2}$ are the effective masses, and $\gamma_{1,2}$ are the Duffing nonlinearities of the two modes. The two modes are found to be dispersively coupled, where the vibration of one mechanical mode creates tension in the beams and in turn modifies the resonance frequency of the other mode (Appendix \ref{section:A}). In Eq.~(\ref{eq1}), the dispersive coupling is represented by the terms with coupling constant $\gamma$ that correspond to coupling energy $\frac{1}{2} \gamma q_1^2 q_2^2$. Parameter $F$ gives the amplitude of the parametric sideband pumping that is proportional to the amplitude of the applied pump current $I_{pump}$. The phase of the pump is $\theta_F$. In the Hamiltonian of the coupled modes, the term corresponding to the pump is $-Fq_1q_2\cos{\left(\omega_Ft+\theta_F\right)}$.

We assume that the pump frequency is close to $\omega_1+ \omega_2$, with the pump detuning $\Delta_F=\omega_F-(\omega_1+\omega_2)$ much smaller than $\omega_F$. By making the substitution $u_1=\frac{1}{2} (q_1-i\omega_1^{-1} {\dot{q}}_1) e^{-i\omega_1 t}$ to convert $q_1$ and $q_2$ to slowly varying complex vibration amplitudes $u_1$ and $u_2$, and by applying the rotating wave approximation, we obtain equations for $u_{1,2}$:
\begin{align}
\label{eq2}
&{\dot{u}}_1+\Gamma_1 u_1-i\frac{3\gamma_{11}}{2\omega_1} u_1 |u_1|^2-i\frac{\gamma_{12}}{\omega_1} u_1 |u_2|^2 \nonumber\\
& = \frac{F_1 e^{i\theta_F}}{4i\omega_1} u_2^\ast, \nonumber\\
&{\dot{u}}_2+(\Gamma_2+i\Delta_F) u_2- i\frac{3\gamma_{22}}{2\omega_2} u_2 |u_2|^2-i\frac{\gamma_{21}}{\omega_2} u_2 |u_1|^2 \nonumber\\
& = \frac{F_2 e^{i\theta_F}}{4i\omega_2} u_1^\ast,
\end{align}
where parameters $\gamma_{11}$ ($1.91\times10^{22}\mathrm{rad^2 s^{-2} m^{-2}}$), $\gamma_{22}$ ($7.38\times10^{25}\mathrm{ rad^2 s^{-2} m^{-2}}$), $\gamma_{12}$ ($8.41\times10^{22}\mathrm{ rad^2 s^{-2} m^{-2}}$), $\gamma_{21}$ ($1.26\times10^{25}\mathrm{ rad^2 s^{-2} m^{-2}}$) and $F_{1,2}$ contain not only direct contributions from $\gamma_1$, $\gamma_2$, $\gamma$ and $F$ respectively, but also account for the renormalization that arises from the nonlinear coupling of the modes \cite{sun_correlated_2016} and from modulating not just the mode coupling, but applying force to one or the other mode \cite{sun_correlated_2016}; we note that $F_1⁄F_2 =\gamma_{12}⁄\gamma_{21} =m_2/m_1$.

We first consider the case in which vibration amplitudes are small so that the terms involving $\gamma_{11}$, $\gamma_{22}$, $\gamma_{12}$ and $\gamma_{21}$, can be neglected. With $\Gamma_2>>\Gamma_1$, we apply the adiabatic approximation by setting ${\dot{u}}_2 = 0$ in the second equation in Eq.~(\ref{eq2}) because after a short transient, both $|{\dot{u}}_1/{u_1}|$ and $|{\dot{u}}_2/{u_2}|$ become much smaller than $\left({\mathrm{\Gamma}_2}^2+\Delta_F^2\right)^{1/2}$. It follows from Eq.~(\ref{eq2}) that for $\Delta_F=0$, $u_2\approx\frac{F_2e^{i\theta_F}}{4i\omega_2\mathrm{\Gamma}_2}u_1^\ast$. Putting $u_2$ back into the first equation yields an effective damping constant of mode 1 that decreases with the pumping power as ${\Gamma_1-\ F}_1F_2/16\omega_1\omega_2\Gamma_2$. 

For sufficiently large pumping power, the effective damping becomes negative and self-sustained vibrations are excited. In fact, both modes start oscillating at the same time, although with different amplitudes. Figures \ref{Fig.1}(d) and \ref{Fig.1}(e) shows these amplitudes for modes 1 and 2 respectively as a function of $\Delta_F$ at pump current amplitude of 189$\ \mu$A. The vibrations are excited when $\Delta_F$ is increased beyond a threshold value $-\omega_c$. This value can be found from the condition of the loss of stability of the stationary state $u_1=u_2=0$, giving \cite{sun_correlated_2016}
\begin{align}
\label{eq3}
&\omega_c=\left(\Gamma_1+\ \Gamma_2\right)\left[\Xi^2-1\right]^{\frac{1}{2}\ },\nonumber\\
&\Xi=\left(F_1F_2/16\omega_1\omega_2\Gamma_1\Gamma_2\right)^{1/2}.
\end{align}
The vibration amplitude increases with $\Delta_F$. At $\Delta_F = \omega_c$, the zero-amplitude state becomes stable again.  Previous experiments \cite{sun_correlated_2016} have determined the dependence of $\omega_c$ on the pump power and studied the bistability and hysteretic behavior as $\Delta_F$ is varied.

%%%%%%%%%%%%%%%%%%%%%%%%%%%%%%%%%%%%%%
%%%%%%%%%%%%%%%%%%%%%%%%%%%%%%

\subsection{Anti-correlated phase diffusion of the two modes}
\label{subsec:2B}

If the pump is not too strong, self-sustained vibrations of the two modes are sinusoidal, but their frequency deviates from $\omega_1$ and $\omega_2$. The deviation depends on the amplitude and frequency of the pump. To analyze the self-sustained vibrations, we write the variables $u_1,u_2$ in the form of the amplitudes $r_1, r_2$ and phases $\phi_1,\phi_2$ of the vibrations and take into account that the vibrations in the rotating frame have frequency $\delta\omega$, 
\begin{align}
\label{eq:change_to_r_phi}
u_1 = r_1 e^{i\phi_1(t)}\exp(i\delta\omega\, t),\quad u_2 = r_2 e^{i\phi_2(t)}\exp(-i\delta\omega\, t).
\end{align}
In the steady state, $r_{1,2}$ and $\phi_{1,2}$ are independent of time. In the laboratory frame, self-sustained oscillations in modes 1 and 2 occur at frequencies of $\omega_\mathrm{self1}=\omega_1 + \delta\omega$ and $\omega_\mathrm{self2}=\omega_2+\Delta_F-\delta\omega$, respectively.

It is convenient to write the equations of motion in the form
\begin{align}
\label{eq:eom_r_phi}
&\dot r_i = -\Gamma_i r_i +  \frac{F_i}{4\omega_i}\frac{r_{3-i}}{r_i}\sin\Theta \quad (i=1,2), \nonumber\\
&\dot\phi_+ = -\Delta_F +  \sum_{i=1,2}\left[\frac{\gamma^{(+)}_{i \,3-i}}{\omega_i}r_{3-i}^2- 
\frac{F_i}{4\omega_i}\frac{r_{3-i}}{r_i}\cos\Theta\right]
\end{align}
and
\begin{align}
\label{eq:eom_phi_-}
&\dot\phi_- = \Delta_F - 2\delta\omega - \sum_{i=1,2}(-1)^i\left[\frac{\gamma^{(-)}_{i \,3-i}}{\omega_i}r_{3-i}^2 - 
\frac{F_i}{4\omega_i}\frac{r_{3-i}}{r_i}\cos\Theta\right],
\end{align}
Here
\[\phi_\pm = \phi_1 \pm \phi_2, \quad \Theta = \theta_F - \phi_+.\]
The explicit form of the coefficients $\gamma^{(\pm)}_{ij}$ is given in Appendix~\ref{section:B}.

The stationary solution of Eqs.~(\ref{eq:eom_r_phi}) and (\ref{eq:eom_phi_-}) gives the amplitudes $r_{1,2}^{(0)}$ and the phases $\phi_{1,2}^{(0)}$ of the self-sustained vibrations of the coupled modes, as well as the vibration frequency in the rotating frame $\delta\omega$. They are also listed in Appendix~\ref{section:B}. In particular, the total phase is 
\begin{align}
\label{eq:phi_stationary}
&\phi_+^{(0)} = \theta_F -\Theta^{(0)}, \quad \Theta^{(0)} = \arcsin (1/\Xi),
\end{align}
where $\Xi$ is defined in Eq.~(\ref{eq3}).

An important feature of Eq.~(\ref{eq:eom_r_phi}) is that the evolution of the amplitudes $r_{1,2}$ and the total phase $\phi_+$ is independent of the phase difference $\phi_-$. At the same time, the evolution of $\phi_-$ depends on $r_{1,2}$ and $\phi_+$ as well as the phase $\theta_F$ of the driving field  but not on $\phi_-$ itself. It is this feature that enables an efficient control of the mode phase.

\begin{figure}
\includegraphics[scale=0.5]{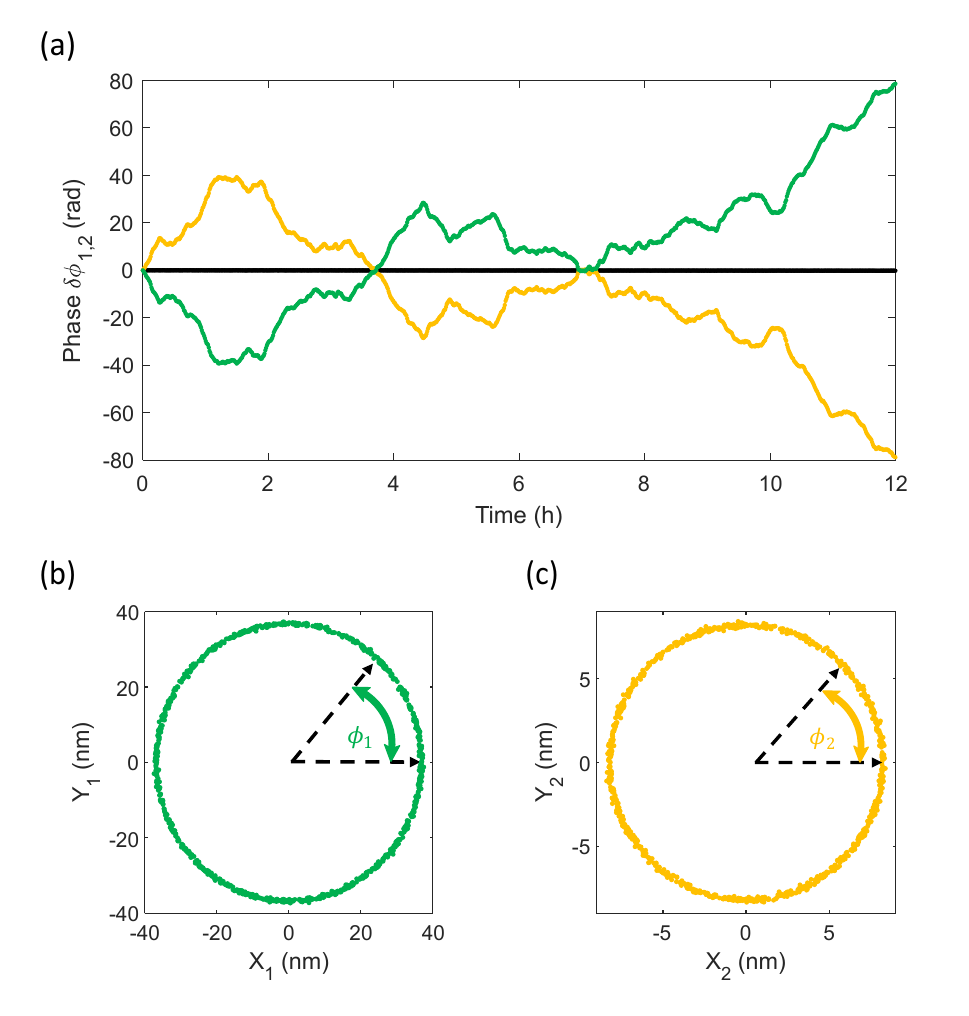}

\caption{ (a) Change of the vibration phase ${\delta\phi}_1$ of mode 1 (green) and ${\delta\phi}_2$ of mode 2 (yellow), measured as a function of time. The black line represents the sum ${\delta\phi}_1+{\delta\phi}_2$. The pump current amplitude is 189 $\mu$A corresponding to $F$ = 2.3 mN$m^{-1}$  and frequency detuning $\Delta_F$ is 300 Hz.  The same results are plotted in panels (b) and (c) in the frames rotating at the frequencies $\omega_\mathrm{self1}$ and $\omega_\mathrm{self2}$ of the self-sustained vibrations of the two modes. The variables are $X_i = r_i \cos(\phi_i)$ and $Y_i = r_i \sin(\phi_i)$, $i = 1, 2$.}

\label{Fig.2} 
\end{figure}

While fluctuations in $\phi_1+\phi_2$ do not accumulate, fluctuations in $\phi_1-\phi_2$ do. Figure \ref{Fig.2}(a) plots the measured phase change ${\delta\phi}_{1,2}(t)= \phi_1(t) - \phi_1(0)$ as a function of time as $\theta_F$ is kept constant. ${\delta\phi}_1(t)$ is almost perfectly anti-correlated with ${\delta\phi}_2(t)$, in agreement with Eq.~(\ref{eq:phi_stationary}). Previous works have demonstrated that the system undergoes phase diffusion, with the mean square phase change increasing with time \cite{sun_correlated_2016}. Depending on the origin of phase fluctuations, the diffusion was found to be either normal or anomalous. This phase diffusion leads to decoherence, determining the linewidth of the self-sustained vibrations. Figures \ref{Fig.2}(b) and (c) plot the results of Fig.~\ref{Fig.2}(a) in frames rotating at the frequencies of self-sustained vibrations of the two modes, $\omega_\mathrm{self1}$ and $\omega_\mathrm{self2}$ respectively. The motion of two modes in the tangential direction is not confined, but is correlated.

%%%%%%%%%%%%%%%%%%%%%%%%%%%%%%%%%%%%%%%%%
%%%%%%%%%%%%%%%%%%%%%%%%%%%%%%%%%%%%%%%%

\section{FREQUENCY STABILIZATION}
\label{sec:three}

With the phase anti-correlation of the two modes, one can measure the phase diffusion of one mode and infer the phase of the other mode without measuring it. This property enables the design of a scheme that uses the measured phase of one mode to stabilize the phase of the other mode, provided that there is a mechanism to alter the latter. We show below that either $\phi_1$ or $\phi_2$ can be stabilized by adjusting the pump phase $\theta_F$ and using the measured $\phi_2$ or $\phi_1$ as the error signal. An antecedent less efficient scheme was implemented in a previous experiment \cite{sun_correlated_2016} to prove the concept for stabilizing mode 2. Here, we analyze in detail, quantitatively,  the effect on the two modes when the pump phase $\theta_F$ is adjusted. Then we develop and optimize the stabilization scheme in our experimental system. Furthermore, we demonstrate that mode 1 can also be stabilized by measuring mode 2, despite requiring much larger changes in $\theta_F$.

%%%%%%%%%%%%%%%%%%%%%%%%%%%%%%%%%%%%%%%%%%%%%%%%%%%

\subsection{Effect of a change of pump phase}
\label{subsec:3A}

For short durations of phase accumulation the noise-induced changes of the phases are  small and can be compensated by small changes of $\theta_F$. The effect can be analyzed by linearizing Eqs.~(\ref{eq:eom_r_phi}) about the stationary solution $r_{1,2}^{(0)}, \phi_+^{(0)}$. It is convenient to write the linearized equations for the increments
\begin{align*}
&x_1 =\zeta - \zeta^{(0)},\quad 
x_2 = \theta_F-\phi_+ - \Theta^{(0)}, \quad x_3 =\frac{r_1-r_1^{(0)}}{r_1^{(0)}},
\end{align*}
where $\zeta \equiv r_2/r_1$. The corresponding equations have the form
\begin{align}
\label{eq:linear_general}
\dot x_n = \Lambda_{nm}x_m \quad (n,m=1,2,3),
\end{align}
where we imply summation over repeated indices. Appendix~\ref{section:B} gives the  explicit form of the matrix $\hat\Lambda$. The initial conditions that corresponds to incrementing $\theta_F$ by $\delta\theta_F$  at $t=0$ are $x_1(0) = x_3(0) = 0, x_2(0) = \delta\theta_F$.

For stable self-sustained vibrations described by Eq.~(\ref{eq:eom_r_phi}) the eigenvalues $\lambda_\nu$ ($\nu=1,2,3$) of the matrix $\hat\Lambda$ have negative real parts. Therefore $x_{1,2,3}(t)$ become exponentially small for $t\gg t_r$, where the relaxation time is $t_r= 1/\min|\mathrm{Re}\, \lambda_\nu|$. This means that, over the relaxation time, $\phi_+$ is incremented by $\delta\theta_F$, i.e., $\delta\phi_+(t)\to \delta\theta_F$. For the rest of the paper, we use $\delta\phi_{\pm}(t)$ and $\Delta\phi_{\pm}$ to denote the transient and settled values of $\phi_{\pm}$ changed from the initial values respectively.   

The eigenvalues $\lambda_k$ can be found analytically near the bifurcation point $\Delta_F = -\omega_c$ where  self-sustained vibrations emerge, see Appendix~\ref{section:B}. Here the eigenvalues $\lambda_{1,2}$ are complex conjugate, with the real part $-(\Gamma_1 + \Gamma_2)$. The eigenvalue $\lambda_3$ is real and small in the absolute value, $\lambda_3 \propto \Delta_F + \omega_c$. Our numerical analysis shows that, for the parameters of our system, the eigenvalue $\lambda_3$ remains small in the absolute value throughout the range we have explored, see Fig.~\ref{Fig.3}~(e) and (f). Therefore the relaxation time is $t_r = - 1/\lambda_3$.

The increment $\delta\phi_-$(t) of $\phi_-$ is described by the linearized equation (\ref{eq:eom_phi_-}), with  $d\delta\phi_-/dt$ being a sum of the appropriately weighted $x_{1,2,3}(t)$. Hence,  $\delta\phi_-$(t) is given by the integrals over time of $x_{1,2,3}(t)$. Since all variables $x_n(t)$ are proportional to $\delta\theta_F$ and all of them decay for $t\gg t_r$, we see that $\delta\phi_-(t)\to C\delta\theta_F$ for $t\gg t_r$, where $C$ is determined by the parameters of the system, see Appendix~\ref{section:C}. Therefore incrementing $\theta_F$ leads, after the transient time $\gtrsim t_r$, to 
\begin{align}
\label{eq:g_parameter}
\Delta\phi_2 =g\delta\theta_F,\quad \Delta\phi_1 = (1-g)\delta\theta_F, \quad g=(1-C)/2.
\end{align}  
%

%%%%%%%%%%%%%%%%%%%%%%%%%%%%%%%%%%%%%%%%%%%%%%%

\subsubsection{Implementing phase increment of the driving field}
\label{subsubsec:delta_Delta}

The increment of the phase $\theta_F$ of the driving field is obtained by changing the driving frequency $\omega_F$ for a short time and then bringing it back to the original value. One scenario is where this time is of order $1/\omega_F$. Equations (\ref{eq2}), (\ref{eq:eom_r_phi}), and (\ref{eq:eom_phi_-}) do not apply on this time scale. However, the change of the ``slow'' variables over time $\sim 1/\omega_F$ associated with the change of $\theta_F$ is small and can be disregarded. It applies to the case implemented in the experiment where $\theta_F$ is controlled by a digital function generator.

Alternatively, the phase can be incremented  by changing the frequency $\omega_F$ over time much longer than $1/\omega_F$ but yet much shorter than $1/\max |\lambda_\nu|$. The dynamics in this case is described by Eqs.~(\ref{eq:eom_r_phi}), and (\ref{eq:eom_phi_-}).  It can be modeled by assuming that  $\Delta_F$ is changed by $\delta\theta_F\delta(t)$. As seen from Eq.~(\ref{eq:eom_r_phi}) this is equivalent to the initial condition $x_2(t\to +0) = \delta\theta_F$. However, $\phi_-$ is also incremented by $\delta\theta_F$.

Yet another way of changing the phases is to apply a pulse of $\Delta_F(t)$ with $\Delta_F(t)$ varying slowly on the scale $t_r$. The change of the variables $x_n, \phi_-$ can be then described in the adiabatic approximation. It leads to simple explicit expressions for this change, see Appendix~\ref{section:C}. Unexpectedly, the result allows one also to express the parameter $C$ in Eq.~(\ref{eq:g_parameter}) in a simple explicit form, which does not require diagonalizing the matrix $\hat\Lambda$. 

We note that the frequency of self-sustained vibrations can be obtained in an alternative way. One could omit $\delta\omega$ in defining the variables $r_{1,2}, \phi_{1,2}$ in  Eq.~(\ref{eq:change_to_r_phi}). This omission will not affect the equations of motion (\ref{eq:eom_r_phi}) for $r_{1,2}, \phi_+$, that is, they will have the same form. The stationary solution of these equations gives $r_{1,2}^{(0)}, \phi_+^{(0)}$, and this solution is stable for $\Delta_F+\omega_c>0$, since Re~$\lambda_{1,2,3}<0$. On the other hand, Eq.~(\ref{eq:eom_phi_-}) will have a solution  $\dot\phi_-$ equals constant. The value of this constant gives $-2\delta\omega$, see Eq.~(\ref{eq:delta_omega}). In the stationary state, the  complex amplitudes $u_{1,2}(t)$ oscillate at frequencies $\delta\omega$ and $-\delta\omega$, respectively.

\subsubsection{Measurement of the effect of a step change of the pump phase}
\label{subsubsec:measure_delta_Delta}

\begin{figure}[t!]
\includegraphics[scale=0.5]{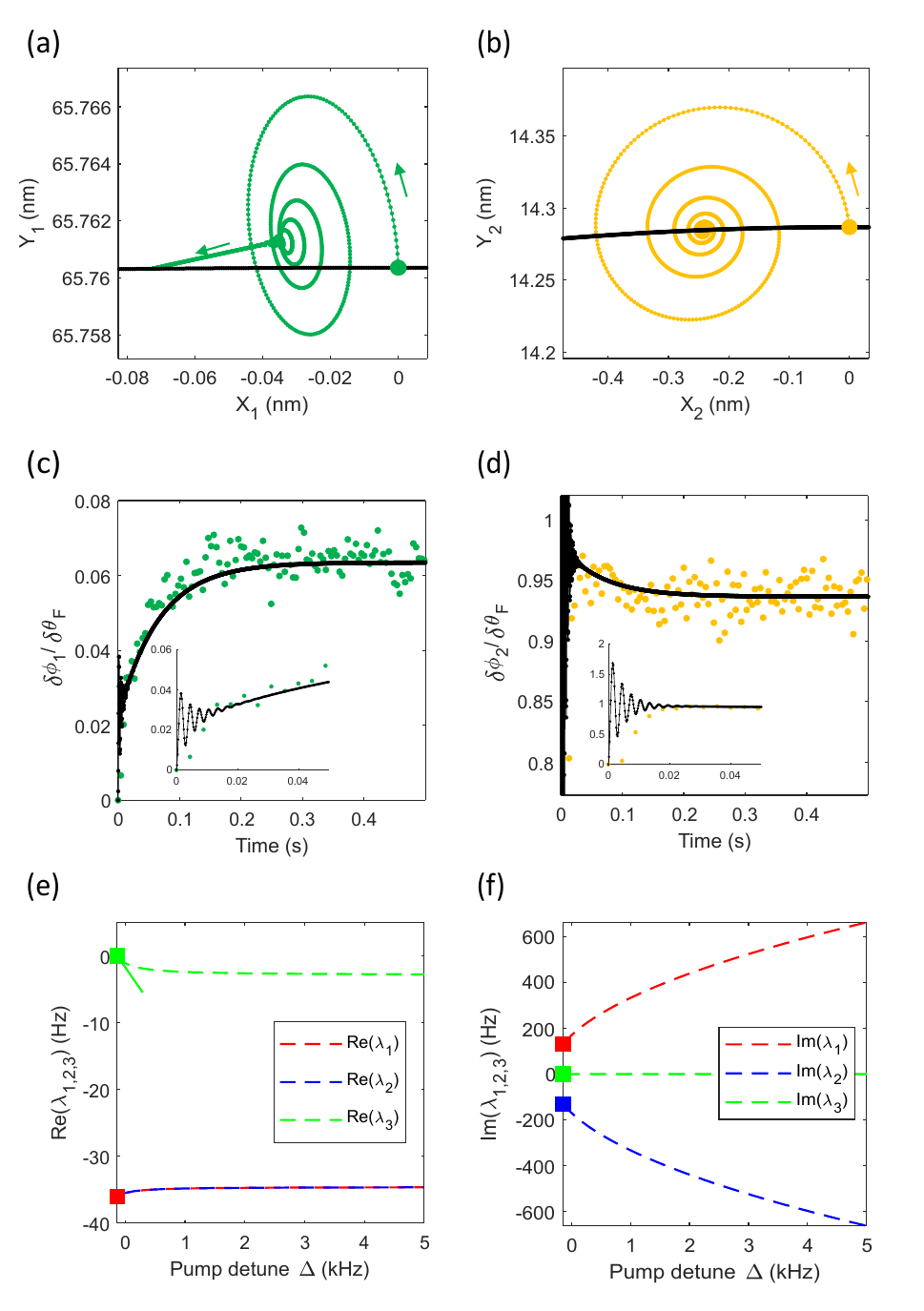}

\caption{(a) Numerical calculation of the dynamics of mode 1 in the frame rotating at frequency $\omega_\mathrm{self1}$ in response to a sudden change of $\theta_F$ by $1^{\circ}$; $X_1 = r_1 \cos \phi_1$ and $Y_1 = r_1 \sin\phi_1$. The pump current amplitude is 379 $\mu$A and frequency detuning $\Delta_F$ is 1000 Hz. Initially, $X_1$ is set to zero. The black line is part of the circles $X_1^2 + Y_1^2 = r_1^{(0)}{}^2$.(b) Corresponding plot for mode 2. (c) Measured dependence of $\phi_1$ on time (green circles). The solid line represents numerical simulations with no fitting parameters. Inset: zoom in at shorter times. (d) Corresponding plot for mode 2. (e) Real parts of the eigenvalues for matrix $\mathbf{\Lambda}$ in the linearized equation of motion plotted as a function of the pump detuned frequency.  $\lambda_{1,2,3}$ are plotted as red, blue and green dashed lines respectively. Near the supercritical Hopf bifurcation point, analytical expressions (Appendix \ref{section:B}) give the green solid line for $\lambda_3$ and the red square for $\lambda_1$. (f) Similar plot for the imaginary parts of $\lambda_{1,2,3}$.}

\label{Fig.3} 
\end{figure}

Figures \ref{Fig.3}(a) and \ref{Fig.3}(b) show that both the amplitude and phase are perturbed by a step change in the pump phase, for modes 1 and 2 respectively. Details of the calculations of the transient are described in Appendix~\ref{section:C}. There are two stages in the transient response. The system first relaxes quickly following the spirals, the decay rate and frequency of which are determined by the real and imaginary parts of the eigenvalues $\lambda_{1,2}$ respectively. As described earlier in this section, $\lambda_{1,2}$ are complex conjugates of each other. The second part of the transient is marked by the arrow pointing to the left in Fig.~\ref{Fig.3}(a) for mode 1. On this part of the phase trajectory the system approaches the new stable state at a much slower rate given by $\lambda_3$\; and the fast variables follow the slow variable adiabatically. For mode 2, the second stage of relaxation occurs over a much smaller range in Fig.~\ref{Fig.3}(b) and is difficult to identify.

After the transient, the amplitudes $r_{1,2}$ of self-sustained vibrations go back to the original values. The phases change by ${\Delta\phi_{1,2}}$ as given by Eq.~(\ref{eq:g_parameter}), assuming that phase diffusion induced by noise is negligible. In particular, ${\Delta\phi}_1+{\Delta\phi}_2$ equals ${\delta\theta}_F$ after the transient, according to Eq.~(\ref{eq:phi_stationary}). 

As described later, the algorithms to stabilize the phases $\phi_{1,2}$ requires knowing the ratio $g={\Delta\phi}_2/\delta\theta_F$  that denotes the fraction of the change in pump phase taken up by mode 2. The fraction of the change in pump phase taken up by mode 1 is $1-g={\Delta\phi}_1/\delta\theta_F$. The value of the parameter $g$ is determined by Eqs.~(\ref{eq:g_parameter}) and (\ref{eq:C_short}). 
We find that $g$ depends only on system parameters including the damping constants, effective masses and nonlinear coupling coefficients, but is independent of the pump detune frequency and amplitude (Appendix \ref{section:C}).  For our system parameters, $g = 0.94$. A change in the phase of the pump therefore affects both mode 1 and mode 2. However, the effect on mode 2 far exceeds that on mode 1.

We measure the response of the two modes for a duration of 0.5 s after $\theta_F$ is abruptly changed by $1^{\circ}$. $\theta_F$ is controlled by a digital function generator. It changes over a time scale $\sim \omega_F^{-1}$, as indicated above, much shorter than $|\lambda_{1,2,3}^{-1}|$. The measurement is repeated 7200 times and the averaged phase changes scaled by the pump step, ${\delta\phi}_{1,2}(t)/\delta\theta_F$, are plotted in Figs.~\ref{Fig.3}(c) and \ref{Fig.3}(d), respectively, as green and yellow dots at detuning $\Delta_F =$ 1000 Hz and amplitude of 379 $\mu$A of the pump current that corresponds to $F$ = 4.5 mN$m^{-1}$. Averaging is necessary to resolve the change from detection noise and the noise introduced by phase diffusion. The agreement between measurement and theory is good on the slow-relaxation time scale, which in our case was 0.3 s, with no fitting parameters. However, it is not feasible for us to measure the fast relaxation represented by the solid lines in the insets, as it requires a much larger bandwidth for the lock-in amplifier that significantly increases the detection noise. 

Apart from the transient response, the measured values of ${\Delta\phi}_{1,2}$ in the steady state also agrees with calculations: mode 2 accounts for most of the phase change (94\%) in response to $\delta\theta_F$ while the remaining 6\% occurs in mode 1. As plotted in Fig.~\ref{Fig.3}(e), the green dashed line represents the eigenvalue $\lambda_3$ that determines the rate of the slow relaxation. $|\lambda_3|$ first increases with increasing pump detuning, then remains largely constant when pump detuning exceeds 2 kHz. The red and blue lines plot the real parts of $\lambda_{1,2}$ that determine the rate of the fast relaxation. Figure \ref{Fig.3}(f) plots the imaginary parts of eigenvalues $\lambda_{1,2}$ that are associated with the frequency of spiraling in Figs.~\ref{Fig.3}(a) and \ref{Fig.3}(b). The relaxation rates change for other pump detuning values. 

The ability to control the phase of the modes by adjusting $\delta\theta_F$, together with the near perfect anti-correlation in the phase diffusion of modes 1 and 2, allows us to implement a scheme to stabilize the phase of mode 2/1 over time scales longer than $|\lambda_3^{-1}|$, by measuring the phase of mode 1/2 that vibrates at a different frequency. 

\section{FREQUENCY STABILIZATION}
\label{sec:four}

\subsection{Stabilization Algorithm}
\label{subsec:4A}

\begin{figure}
\includegraphics[scale=0.5]{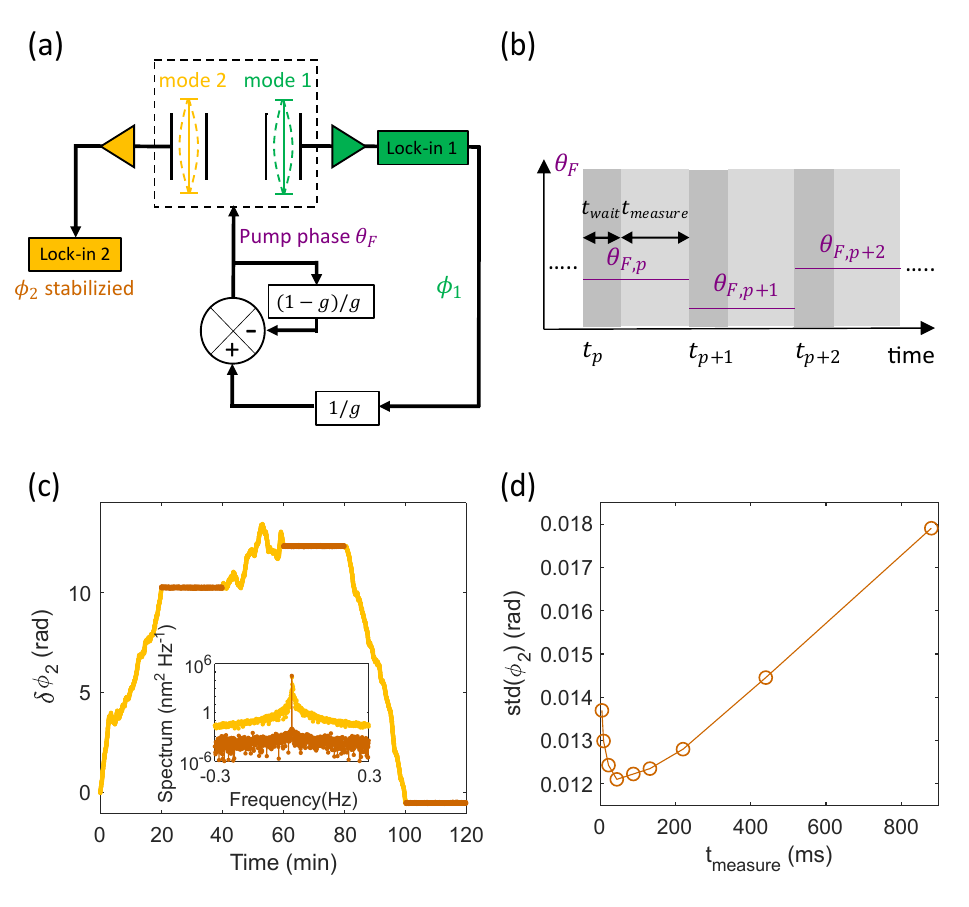}

\caption{(a) Algorithm for stabilizing $\phi_2$ by measuring $\phi_1$ and adjusting $\theta_F$. (b) The duration of one cycle equals $t_\mathrm{wait} + t_\mathrm{measure}$. The $p^{th}$ cycle lasts from time $t_p$ to $t_{p+1}$. $\theta_F$ is changed only at the beginning of each cycle.  (c) $\phi_2$ is measured as a function of time when the stabilization routine is turned off for 20 min (yellow) and on for 20 min (brown). The pump current amplitude is 189 $\mu$A and frequency detuning is 300 Hz.  Inset: The spectra of the self-sustained vibrations of mode 2  when the stabilization algorithm in (b) is implemented. The spectral width is significantly reduced with stabilization turned on. (d) Dependence of the standard deviation of $\phi_2$ on $t_\mathrm{measure}$ at a fixed $t_\mathrm{wait}$ of 300 ms.}

\label{Fig.4} 
\end{figure}

We first discuss how the phase of mode 2 can be stabilized by measuring mode 1. Figure \ref{Fig.4}(a) shows the stabilization algorithm. The procedure starts at time t = 0, at which we define $\phi_1\left(t=0\right)=\phi_2\left(t=0\right)=\ \ \theta_F\left(t=0\right)=0$ for convenience. In each cycle, $\phi_1$ is measured. Then the pump phase $\theta_F$ is adjusted accordingly in the subsequent cycle to bring $\phi_2$ close to zero. The duration of each cycle $\Delta t$ is chosen to exceed the relaxation time $|\lambda_3^{-1}|$. The $p^{th}$ cycle lies in the time interval $t_p<t<t_{p+1}$, where $t_p=p\Delta t$ and $p$ is a positive integer. $\theta_F$ is modified only at the beginning of a cycle at $t_p$, to a value $\theta_{F,p}$ determined by the algorithm described below. It remains fixed at other times. 

Let $\phi_{1,\mathrm{diffuse}}\left(t\right)$ and $\phi_{2,\mathrm{diffuse}}\left(t\right)$ represent the phase change of modes 1 and 2 respectively that arises solely due to phase diffusion. In the $p^{th}$ cycle, after the system has settled in response to changes in $\theta_F$ that occurred at $t_p$, the phases of the two modes are given by:
\begin{align}
\label{eq15}
&\phi_1\left(t\right)=\phi_{1,\mathrm{diffuse}}\left(t\right)+\left(1-g\right)\theta_{F,p},\nonumber\\
&\phi_2\left(t\right)=\phi_{2,\mathrm{diffuse}}\left(t\right)+g\theta_{F,p}.
\end{align}
Here, the second terms on the right-hand side account for the effect of the change in pump phase $\theta_{F,p}$ that is split between modes 1 and 2 according to a fixed ratio $g$ that was discussed in section \ref{subsec:3A}. Our goal is to choose $\theta_{F,p+1}$ in the next cycle to maintain $\phi_2\left(t\right)$ close to zero. As discussed in section \ref{subsec:2B}, there is near perfect anti-correlation in the phase diffusion of modes 1 and 2 for fixed $\theta_{F,p}$. In other words, $\phi_{2,\mathrm{diffuse}}\left(t\right)$ can be inferred by measuring mode 1, without measuring mode 2 itself. Rearranging the first equation in Eq.~(\ref{eq15}) yields:
\begin{align}
\label{eq16}
&\phi_{2,\mathrm{diffuse}}\left(t\right)\ =-\phi_{1,\mathrm{diffuse}}\left(t\right)\ = \left(1-g\right)\theta_{F,p}-\phi_1\left(t\right).
\end{align}
The measurement of $\phi_1\left(t\right)$, however, takes a finite amount of time of data averaging to reduce the noise introduced by the detection circuit. As shown in Fig.~\ref{Fig.4}(b), at the beginning of each cycle, the system is first allowed to settle in response to the change in $\theta_F$. Then the measurement of $\phi_1$ is performed over time $t_\mathrm{measure}$ to yield an averaged value ${\widetilde{\phi}}_{1,p}$. The pump phase $\theta_{F,p+1}$ for the next cycle is chosen by replacing $\phi_1\left(t\right)$ in Eq.~(\ref{eq16}) by ${\widetilde{\phi}}_{1,p}$ and then substituting into the second equation in Eq.~(\ref{eq15}) with $\phi_2\left(t\right)$ set to zero, yielding:
\begin{align}
\label{eq17}
&\theta_{F,p+1}=\frac{1}{g}[{\widetilde{\phi}}_{1,p}-\left(1-g\right)\theta_{F,p}].
\end{align}
For stabilization of the phase of mode 2, the pump phase for the $(p+1)^{th}$ cycle is thus determined by both the measured phase of mode 1 and the pump phase chosen for the $p^{th}$ cycle. 

To keep $\phi_2(t)$ small we choose the time interval $\Delta t$ to be short. Consequently, the  increment $\theta_{F,p+1} - \theta_{F,p} = (\tilde \phi_{1,p}-\theta_{F,p})/g$ is small. To understand the evolution of the increments of $\theta_{F,p}$ we assume that right before the $(p+1)$th correction, i.e., for $t=t_{p+1}-\delta$ with small $\delta>0$, there are small deviations of $\phi_2$ from the target value of zero, so that $\phi_2(t) =\ep_p$, with $|\ep_p|\ll 1$ being a random number.  If we set  $\phi_1(t_{p+1}-\delta) = \tilde\phi_{1,p}$, i.e., if we disregard the change of  $\phi_1(t)$ over time $  t_\mathrm{measure}$   and ignore the uncertainty in measuring $\phi_1$,  we have  from Eq.~(\ref{eq15})  $\tilde \phi_{1,p}= \theta_{F,p} - \ep_p$. The increment in pump phase is then given by: 
\begin{align}
\label{eq17a}
&\theta_{F,p+1} - \theta_{F,p}= -\ep_p/g.
\end{align}
With $g \sim 0.94$ in our experiment, $|\theta_{F,p+1} - \theta_{F,p}|$ is indeed small. In a way one can think of this equation  as a discrete analog of a Langevin equation. The values of $\ep_p$ are determined by the imprecision of the measurements of the phase and  of the control, and it is reasonable to assume that $\ep_p$ with different $p$ are incorrelated. Then $\braket{\theta_{F,p}^2}$ increases with the number of cycles as $p\braket {\ep_p^2}/g^2$.

\begin{figure}
\includegraphics[scale=0.5]{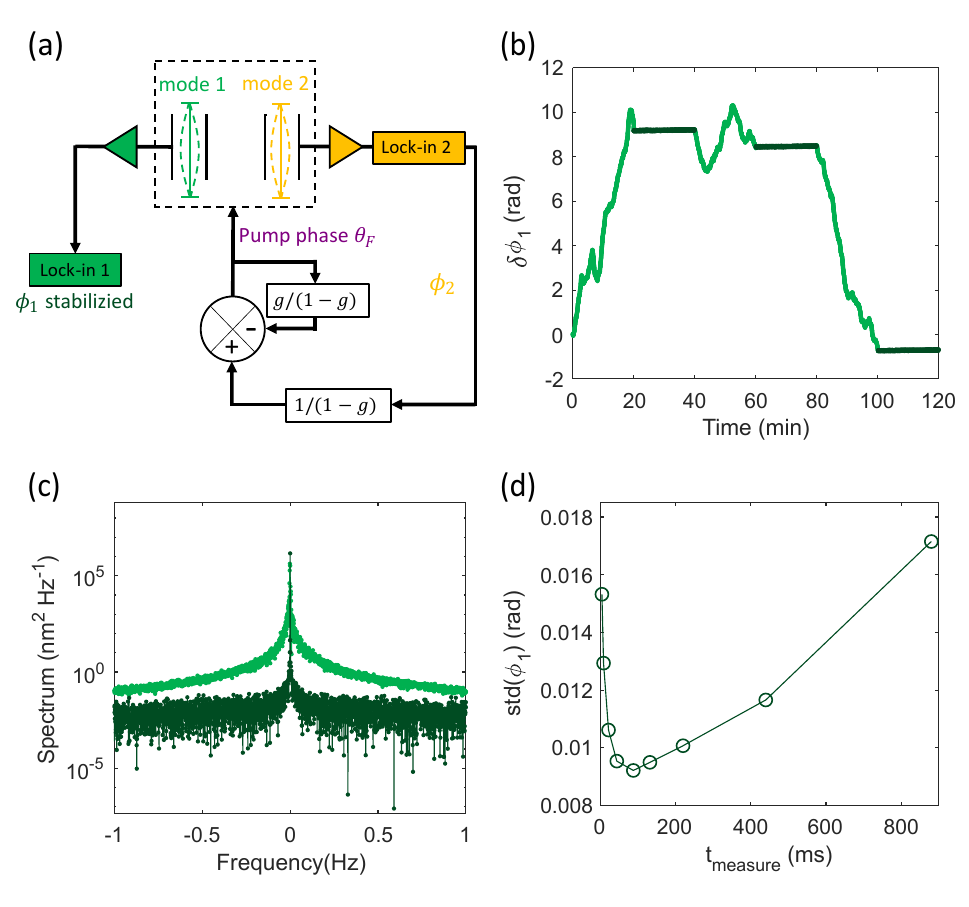}

\caption{(a) Algorithm for stabilizing $\phi_1$ by measuring $\phi_2$ and adjusting $\theta_F$. (b) $\phi_1$ is measured as a function of time when the stabilization routine is turned off (green) and on (dark green). The pump current amplitude is 189 $\mu$A and frequency detune is 300 Hz. (c) The spectra of the self-sustained vibrations of mode 1 with and without stabilization. (d) Dependence of the standard deviation of $\phi_1$ on $t_\mathrm{measure}$ at a fixed $t_\mathrm{wait}$ of 300 ms.}

\label{Fig.5} 
\end{figure}

As discussed in section \ref{subsec:3A}, a step change in $\theta_F$ leads to changes in both $\phi_1$ and $\phi_2$. Therefore the phase of mode 1 can also be stabilized by measuring the phase of mode 2 and adjusting $\theta_F$ to compensate for the phase diffusion. However, since $\frac{{\Delta\phi}_1}{\delta\theta_F}=1-g << 1$, the required $\delta\theta_F$ is much larger when compared to what is needed for stabilization of mode 2, for comparable phase diffusion. The algorithm for stabilizing mode 1 is shown in Fig.~\ref{Fig.5}(a). It is similar to that for stabilizing mode 2, except that mode 2 is measured instead and the factor $g$ in Eq.~(\ref{eq17})  and Eq.~(\ref{eq17a}) is replaced with $1-g$:
\begin{align}
\label{eq18}
&\theta_{F,p+1}=\frac{1}{1-g}\left[{\widetilde{\phi}}_{2,p}-g\theta_{F,p}\right]
\end{align}
\begin{align}
\label{eq18a}
&\theta_{F,p+1} - \theta_{F,p}= -\ep_p/(1-g),
\end{align}
where ${\widetilde{\phi}}_{2,p}$ is the measured value of $\phi_2$ at the end of the $p^{th}$ cycle. In Eq.~(\ref{eq18}), $\theta_{F,p}$ is multiplied by $\sim$ 16. It is therefore essential to take into account the effect of the pump phase change on mode 2 for the stabilization scheme to work properly for mode 1. By comparing Eq.~(\ref{eq18a}) to Eq.~(\ref{eq17a}), we find that the increment in pump phase required for stabilizing mode 1 is much larger than that for stabilizing mode 2, which is consistent with the fact that mode 1 only takes up a small fraction ($\sim 6\%$) of a step change of the pump phase.

\subsection{Implementation}
\label{subsec:4B}

We demonstrate stabilization of the vibration frequencies for a pumping current of 189 $\mu$A at detune frequency $\Delta_F = 300$ Hz. Figure \ref{Fig.4}(c) shows how the phase of mode 2 is stabilized by adjusting $\theta_F$ based on measurement of $\phi_1. \ \phi_{1,2}$ are measured with two lockin amplifiers referenced to $\omega_\mathrm{self1,2}$ respectively, with time constant of 1 ms. Measurements are recorded every 4.4 ms. $t_\mathrm{wait}$ is chosen as 300 ms that exceeds the settling time in response to a step change in $\theta_F$. $t_\mathrm{measure}$ is chosen as 44 ms. Without stabilization, $\phi_2$ undergoes diffusion (yellow). The phase fluctuations determine the width of the spectral peak in the inset of Fig.~\ref{Fig.4}b. When the stabilization is turned on (brown), fluctuations in $\phi_2$ at time scales exceeding $\sim 1$ s are significantly reduced. Spectral components close to $\omega_\mathrm{self2}$ are suppressed to levels given by the noise in the detection circuit. For example, the single side-band phase noise at offset of 0.1 Hz is reduced by 29 dB with phase stabilization turned on.

We optimize the stabilization algorithm with respect to $t_\mathrm{measure}$. Figure \ref{Fig.4}(d) plots the standard deviation $\sigma_{\phi_2}$ i.e., the root mean squared value of $\phi_2$, measured as a function of $t_\mathrm{measure}$ for constant $t_\mathrm{wait}$ of 300 ms. As $t_\mathrm{measure}$ increases from the minimum value of 4.4 ms, $\sigma_{\phi_2}$ initially decreases because a longer $t_\mathrm{measure}$ reduces the uncertainty introduced by the detection noise in measuring $\phi_1$ that is used to determine $\theta_F$ in the subsequent stabilization cycle. However, with further increase of  $t_\mathrm{measure}$, $\sigma_{\phi_2}$ increases. This  rise occurs because the value $\tilde\phi_{1p}$, which is used to determine the control signal $\theta_{F,p+1}$ and is obtained by averaging $\phi_1(t)$ over time $t_\mathrm{measure}$, is increasingly different from the actual value $\phi_1(t_{p+1})$  required for efficient control. This difference is due to the phase diffusion. The optimal $t_\mathrm{measure}$ depends on  the detection noise and both the intensity and the correlation time of the noise that leads to the phase diffusion. It should be shorter than this correlation time. We find that in our experiment, when 
$t_\mathrm{measure} > 200$ ms, phase diffusion becomes the dominant effect and leads to the rise in $\sigma_{\phi_2}$. The optimal measured $\sigma_{\phi_2}$ is $\sim 12$ mrad attained at $t_\mathrm{measure}$ $\sim 44$ ms. 

Figure \ref{Fig.5}(b) shows the stabilization of the phase of mode 1 by adjusting $\theta_F$ based on measurement of $\phi_2$ using Eq.~(\ref{eq18}). The spectral width of mode 1 is also reduced significantly, as shown in Fig.~\ref{Fig.5}(c). In Fig.~\ref{Fig.5}(d), the optimal value of $\sigma_{\phi_1}$ is $\sim 9.5$ mrad attained at $t_\mathrm{measure}\sim 88$ ms. A larger $t_\mathrm{measure}$ is needed compared to the case of stabilizing mode 2, consistent with the fact that the uncertainty in detecting $\phi_2$ is larger than that of $\phi_1$ in our experiment.

\section{DISCUSSION AND OUTLOOK}
\label{sec:five}

Efficient stabilization of the phase of one mode relies on accurate determination of the phase of the other mode. The effect of the noise introduced by the detection circuits on the latter can be reduced by increasing the amplitude of the self-sustained oscillations. One way to achieve high amplitudes for the self-sustained vibrations is to choose large pump detuning, as shown in Fig.~\ref{Fig.1}(d) and \ref{Fig.1}(e). When the pump detuning is chosen to exceed $\omega_c$, the zero-amplitude state becomes stable. For parameters chosen in our experiment, we have not encountered problems with stabilizing mode 2. However, in stabilizing mode 1, the change in the pump phase required to compensate for the phase diffusion is much larger than the phase diffusion itself [Eq.~(\ref{eq18})]. We find that occasionally the pump phase changes of up to 20$^{\circ}$ are required in our experiment. Such large change of the pump phase could perturb the system to switch to the zero-amplitude state if the pump detuning is large. To avoid this problem, it is necessary to limit the pump detuning frequency to values smaller than $\omega_c$ so that the zero amplitude state does not become stable.

In the stabilization algorithm, the pump phase is adjusted once per cycle. As a result, phase diffusion at time scales shorter than the cycle period are not compensated for. The duration of one stabilization cycle $\Delta t$ is the sum of $t_\mathrm{wait}$ and $t_\mathrm{measure}$. In our experiment, $t_\mathrm{wait}$ is chosen to barely exceed the settling time of the system in response to in step change in the pump phase. As shown in Figs. \ref{Fig.4}(d) and \ref{Fig.5}(d), $t_\mathrm{measure}$ that gives optimal $\sigma_{\phi_{1,2}}$ is much shorter than $t_\mathrm{wait}$. Further improvements to extend the stabilization to shorter time scales would rely on tuning system parameters (Appendix \ref{section:E}) or design new resonators to achieve smaller settling time. 

To summarize, we present an algorithm to stabilize the frequency of self-sustained oscillations that are induced by dynamical backaction in a sideband-driven micromechanical resonator. It is based on periodically  shifting the phase of the pump to compensate for phase diffusion. When the system undergoes phase diffusion in the presence of weak noise, the phases of modes 1 and 2 add up to a constant (within our detection limit) that can be adjusted by the phase of the sideband pump. For a step change of the pump phase, the phases of both modes settle to new values after a transient. The duration of the transient is determined by the smallest real part of the eigenvalues of the linearized equations of motion about the vibrational state. If the pump change is small, the phase change of each mode is proportional to the pump phase change. The two proportionality constants add up to one.  This finding, together with the phase anti-correlation of the two modes, allow us to stabilize the phase of one mode by measuring the phase of the other mode and then compensating for the phase diffusion by adjusting the phase of the pump. We demonstrate that phase fluctuations of either the high or low frequency mode can be significantly reduced, resulting in a much narrower spectral linewidth.  The control algorithm requires knowledge of the phase shifts of the modes in response to a shift in the pump phase. After the transient, the proportionality constant   between the mode phase shifts and the pump phase shift is independent of the parameters of the pump. It is fully determined by the parameters of the resonators.

The results open new opportunities in generating stable mechanical vibrations via parametric down-conversion in nonlinear micromechanical resonators, with controlled amplitude and phase. We reiterate that our scheme is distinct from direct feedback that involves stabilizing a particular mode by measuring its phase because a frequency reference near this mode is not required.  As long as a frequency reference is available to determine the phase diffusion of one of the modes, the phase of the other mode can be stabilized even if its eigenfrequency is orders of magnitude different from that of the former. In principle, one can design resonators with the upper mode reaching microwave frequencies in which the construction of signal sources with low phase noise and small long term drift requires complex circuitry. The analysis also can be extended to mechanical resonators coupled to optical or microwave cavities \cite{kippenberg_cavity_2008,aspelmeyer_cavity_2014}.

%%%%%%%%%%%%%%%%%%%%%%%%%%%%%%%%
%%%%%%%%%%%%%%%%%%%%

{\it Acknowledgements:}
This work is supported by the Research Grants Council of Hong Kong SAR, China (Project No. 16304219). M.I.D. acknowledges partial support  from the US Defense Advanced Research Projects Agency (Grant No. HR0011-23-2-004) and from the Moore Foundation (Grant No. 12214)
 
 %%%%%%%%%%%%%%%%%%%%%%%%%%%%%%%%

%\begin{appendix}
\appendix
\section{DISPERSIVE COUPLING BETWEEN MODES 1 AND 2}
\label{section:A}

\begin{figure}
\includegraphics[scale=0.5]{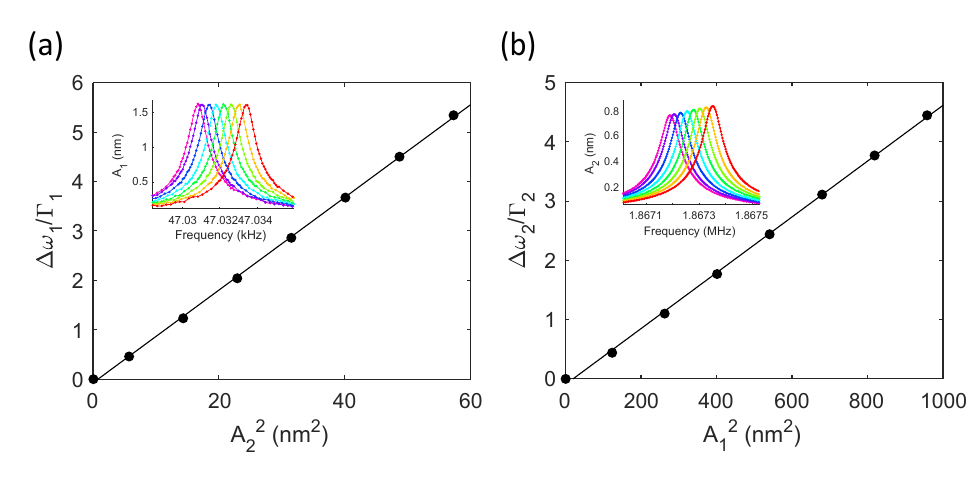}

\caption{(a) Scaled shifts of the frequency of mode 1 versus the square of the amplitude $A_2$ of forced vibrations in mode 2. The line is a linear fit through the origin. Inset: Spectra of mode 1 under a periodic electrostatic force. Each resonance peak corresponds to a data point in the main panel. (b) A similar plot for mode 2.}

\label{Fig.6} 
\end{figure}

Modes 1 and 2 in the resonator are dispersively coupled, with interaction energy of the form $\frac{1}{2}\gamma q_1^2q_2^2$. Figure \ref{Fig.6} shows that the decrease in the resonant frequency of one mode is proportional to the square vibration amplitude of the other mode. In this measurement, the sideband pumping is removed. Instead, ac voltages at frequencies close to $\omega_{1,2}$ are applied to the side gates of modes 1 and 2 respectively to generate periodic electrostatic forces to excite forced vibrations at amplitudes $A_{1,2}$. From the slopes of the linear fits, $\gamma$ is determined to be $6.41\times10^{12} \ kg\ rad^2s^{-2}m^{-2}$.

%%%%%%%%%%%%%%%%%%%%%%%%%%%%%%%%%%%%%%%%%
%%%%%%%%%%%%%%%%%%%%%%%%%%%%%%%%%%%%%%%%%%

\section{EIGENVALUES FOR THE LINEARIZED EQUATIONS OF MOTION NEAR THE SUPERCRITICAL BIFURCATION POINT}
\label{section:B}

The nonlinearity parameters in the equation for the phases of self-sustained vibrations (\ref{eq:eom_r_phi}) have the form
\begin{align}
\label{eq:gamma_pm}
 \gamma^{(\pm)}_{ij}=\gamma_{ij}\pm \frac{3\gamma_{jj}\omega_i}{2\omega_j}  \quad (i,j=1,2; i\neq j)
 \end{align}

By setting $\dot r_{1,2} = \dot\phi_\pm = 0$ in Eqs.~(\ref{eq:eom_r_phi}), we obtain the ratio of the stationary amplitudes $r_{1,2}^{(0)}$ and the phase  $\phi_+^{(0)} \equiv \phi_1^{(0)} + \phi_2^{(0)}$, cf. Ref.~\cite{sun_correlated_2016}:
\begin{align}
&\sin\Theta^{(0)}=1/\Xi,\quad  \Theta^{(0)}= \theta_F - \phi_+^{(0)},\nonumber\\ 
&\zeta^{(0)}\equiv r_2^{(0)}/r_1^{(0)}=(\Gamma_1F_2\omega_1/\Gamma_2F_1\omega_2)^{1/2},
\end{align}
where $\Xi$ is given by Eq.~(\ref{eq3}).

The vibration frequency in the rotating frame $\delta\omega$ and the squared radius of the vibrations of mode 1 are given by the expressions
\begin{align}
\label{eq:delta_omega}
&\delta\omega=\left(\frac{\gamma_{12}}{\omega_1}\zeta^{(0)}{}^2+\frac{3\gamma_{11}}{2\omega_1}\right){r_1^{(0)}}^2-\Gamma_1\left(\Xi^2-1\right)^\frac{1}{2}\nonumber\\
&{r_1^{(0)}}^2=\frac{\Gamma_2\omega_2F_1}{\gamma^{(+)}_{21}\Gamma_2F_1+\gamma^{(+)}_{12}\Gamma_1F_2}\left(\Delta_F+\omega_c\right)
,
\end{align}
where the critical detuning $\omega_c$ is given by Eq.~(\ref{eq3}), $\omega_c = (\Gamma_1 + \Gamma_2)(\Xi^2 - 1)^{1/2}$.

The linearized equations of motion (\ref{eq:linear_general}) 
 for the increments 
\[x_1 = \delta\zeta,\quad  x_2 = \theta_F - \delta\phi_+ - \Theta^{(0)},\quad x_3 = \delta r_1/r_1^{(0)}\]
of the dynamical variables $\zeta = r_2/r_1, \phi_+ = \phi_1 + \phi_2$, and $r_1$  follow from Eq.~(\ref{eq:eom_r_phi}). We will write them in a vector form for the vector $\xb$ with components $x_1,x_2,x_3$. We emphasize that the unknown quantity is $\delta\phi_+$.  It is a dynamical variable, whereas $\theta_F$ is the control parameter. Therefore $\dot x_2 \equiv -d\delta\phi_+/dt$.  The equations read

\begin{widetext}
\begin{align}
\label{eq:linearized_zeta_phi}
&\dot\xb = \hat\Lambda \xb;\quad \Lambda_{11} = -\left(\Gamma_1+\Gamma_2\right), \quad 
\Lambda_{12} = \left(\Gamma_2-\Gamma_1\right)\zeta^{\left(0\right)}\left(\Xi^2-1\right)^{1/2}, \quad \Lambda_{13} = 0\nonumber\\
&\Lambda_{21} =  -\left(\Gamma_2-\Gamma_1\right)\left(\Xi^2-1\right)^{1/2}/\zeta^{(0)} 
-\left(2\gamma_{12}^+/\omega_1 \right){r_1^{\left(0\right)}}^2\zeta^{\left(0\right)},\quad
\Lambda_{22} = -(\Gamma_1 + \Gamma_2), \quad 
\Lambda_{23} = -2[\left(\gamma_{12}^+/\omega_1\right){\zeta^{\left(0\right)}}^2+(\gamma_{21}^+/\omega_2)]r_1^{\left(0\right)}{}^2,\nonumber\\
&\Lambda_{31} = \left(\Gamma_1\Gamma_2F_1\omega_2/F_2 \omega_1\right)^{1/2} %r_1^{\left(0\right)},
,\quad \Lambda_{32} = \Gamma_1\left(\Xi^2-1\right)^{1/2},%r_1^{\left(0\right)}, 
\quad \Lambda_{33}=0.
\end{align}

\end{widetext}

The analysis of the dynamics is simplified near the supercritical bifurcation point $\Delta_F = -\omega_c$ where the state of self-sustained vibrations emerges. Here $r_{1,2}^{(0)}<<1$, as seen from Eq.~(\ref{eq:delta_omega}), whereas the ratio $\zeta^{(0)}$ is not small. Therefore, as a first step, we can set $r_1^{(0)}=0$ in Eqs.~(\ref{eq:linearized_zeta_phi}). This leads to a system of two coupled equations for $\delta\zeta$ and $\delta\phi_+$. The corresponding eigenvalues are 
\begin{align}
\label{eq:eigenvalues_1_2_critical}
\lambda_{1,2} \approx -\left(\Gamma_1+ \Gamma_2\right)\pm i \omega_c(\Gamma_2-\Gamma_1)/(\Gamma_2+\ \Gamma_1).
\end{align}
Both the real and the imaginary parts of $\lambda_{1,2}$ are finite at the bifurcation point, as shown in Figs.~\ref{Fig.3}(e) and (f). 

The third eigenvalue of the matrix $\hat \Lambda$ is small, $|\lambda_3| \ll \Gamma_{1,2}$ near the bifurcation point. It describes the slow decay of $x_3\equiv\delta r_1$. The variables $x_1, x_2$ follow this decay adiabatically. They can be found from Eq.~(\ref{eq:linearized_zeta_phi}) by setting  $\dot x_1 = \dot x_2 = 0$. Then both $x_1$ and $x_2$ are proportional to $\Lambda_{23}\propto r_1^{(0)}{}^2x_3$. The equation for $x_3$ then takes the form $\dot x_3 =\lambda_3 x_3$ with $\lambda_3 \propto - r_1^{(0)}{}^2 \propto \Delta_F + \omega_c$. 

%

%\section{DEPENDENCE OF THE SUM OF PHASE OF THE TWO MODES ON THE PUMP PHASE}
%\label{section:C_old}
%\HBC{HBC: Mark suggests deleting this section. It is not converted into latex.}

%%%%%%%%%%%%%%%%%%%%%%%%%%%%%%%%%%%%%%%%%%
%%%%%%%%%%%%%%%%%%%%%%%%%%%%%%%%%%%%%%%%%%%

\section{INCREMENT OF THE PHASE DIFFERENCE}
\label{section:C}

In this section, we find the values of the phases $\phi_1,\phi_2$ after the transient process that follows incrementing the driving field phase $\theta_F$ by $\delta\theta_F$. This can be done by diagonalizing the matrix $\hat \Lambda$. Respectively, we introduce right and left eigenvectors of this matrix, $\yb_\nu$ and $\bar \yb_\mu$,
\[\hat\Lambda \yb_\nu = \lambda_\nu \yb_\nu,\quad   \bar\yb_\nu\hat\Lambda = \lambda_\nu\bar\yb_\nu, 
\quad \bar\yb_{\nu'}\cdot\yb_\nu = \delta_{\nu\nu'}. \]
We then expand $\xb(t)$ in $\yb$ and use Eq.~(\ref{eq:linearized_zeta_phi}) to obtain
\begin{align}
\label{eq:lin_algebra}
&\xb(t) = \sum_\nu C_\nu(t)\yb_\nu, \quad  C_\nu(t) = \exp(\lambda_\nu t)C_\nu(0),\nonumber\\
&C_\nu(0) = \bar\yb_\nu \cdot \xb(0)
\end{align}

To fully describe time evolution of the coupled oscillators in response to a perturbation, and in particular in response to the increment of the phase of the drive $\theta_F$, this equation has to be complemented by the expression for the increment of the phase $\delta\phi_-$. The equation for $\delta\phi_-$ can be obtained by linearizing Eq.~(\ref{eq:eom_phi_-}), which gives
\begin{align}
\label{eq:phi_-_linearized}
&\frac{d}{dt} \delta\phi_- = %\delta\theta_F\delta(t) +
{\bm \Phi}\cdot\xb, \quad \Phi_1 = -(\Gamma_1 + \Gamma_2)(\Xi^2 -1)^{1/2}/\zeta^{(0)} 
\nonumber\\
&+ 2(\gamma_{12}^-/\omega_1)r_1^{(0)}{}^2 \zeta^{(0)}, \quad \Phi_2 = \Gamma_1 - \Gamma_2, \nonumber\\ 
& \Phi_3 = 2[(\gamma_{12}^-/\omega_1)\zeta^{(0)}{}^2-  (\gamma_{21}^-/\omega_2)] r_1^{(0)}{}^2.
\end{align} 
The solution of Eqs.~(\ref{eq:lin_algebra}) and (\ref{eq:phi_-_linearized}) was used in Fig.~\ref{Fig.3} to illustrate the evolution of the parameters in time. 

For $t\gg t_r$, where $t_r=\max |\mathrm{Re}\, \lambda_{1,2,3}|^{-1}$, all components of the vector $\xb(t)$ go to zero, and therefore $\delta\phi_+(t)\to \delta\theta_F$. To find the change of $\phi_-$ we integrate Eq.~(\ref{eq:phi_-_linearized}) over time, which gives in the limit of large time
\begin{align}
\label{eq:phi_-_asymptote}
\delta\phi_-(t) \to - \sum_\nu \bigl(\bar\yb_\nu\cdot \xb(0)\bigr) \bigl({\bm\Phi}\cdot \yb_\nu\bigr)/\lambda_\nu.
%\sum_{n,m}\Phi_n U_{nm} (U^{-1})_{m2}\lambda_m^{-2}\delta\theta_F
\end{align}
For $\xb(0) = (0,\delta\theta_F,0)$ this gives $\delta\phi_-(t)\to C\delta\theta_F$, with 
\begin{align}
\label{eq:C_parameter}
C=-\sum_\nu (\bar\yb_\nu)_2 \bigl({\bm\Phi}\cdot \yb_\nu\bigr)/\lambda_\nu.
\end{align}
This expression has been used in Eq.~(\ref{eq:g_parameter}) to find the values approached by the phases $\phi_1$ and $\phi_2$ after a change of the phase of the drive. We note that, even though two of the three eigenvalues $\lambda_\nu$ are complex, it is clear that $C$ is real. 

%%%%%%%%%%%%%%%%%%%%%%%%%%%%%%%%%%%%%%%%%

\subsection{Adiabatic change of the drive frequency}
\label{subsec:adiabatic}

An insight into the value of the parameter $C$ can be gained by studying the dynamics of the system for slowly varying frequency of the driving field, where $\Delta_F$ slowly increases and then decreases back to its value, $\Delta_F \to \Delta_F + \delta\Delta_F (t)$. We assume that $\delta\Delta_F (t)$ is small and $|d\delta\Delta_F (t)/dt|\ll |\delta\Delta_F (t)|/t_r$. In this case, as seen from Eq.~(\ref{eq:eom_r_phi}),  the equation of motion for the increments $\xb$ takes the form
\[ \dot x_i = \sum_j\Lambda_{ij} x_j  -\delta\Delta_F (t)\,\delta_{i,2}, \quad \xb(0) = {\bf 0}.\]
Changing to variables $\yb$, as in Eq.~(\ref{eq:lin_algebra}), we obtain for the expansion coefficients $C_\nu$   
\begin{align}
\label{eq:adiabat_C_nu}
\dot C_\nu = \lambda_\nu C_\nu -\delta\Delta_F (t)(\bar\yb_\nu)_2.
\end{align}
where $(\bar\yb_\nu)_2$ is the second component of the vector $\bar\yb_\nu$. 

Assuming that $\delta\Delta_F (t)$ varies slowly, we can disregard $\dot C_\nu$ in this equation, which gives  
\[C_\nu = \lambda_\nu^{-1} (\bar\yb_\nu)_2 \delta\Delta_F (t).\]
After a slow pulse $\delta\Delta_F (t)$ the coefficients $C_\nu$ become equal to zero.

We now consider Eq.~(\ref{eq:eom_phi_-}) for $\delta\phi_-$. It has the form
\begin{align}
\label{eq:adiabat_phi_-}
&\frac{d}{dt}\delta\phi_- = \delta\Delta_F (t) \sum_\nu (\bar\yb_\nu)_2 \bigl({\bm\Phi}\cdot \yb_\nu\bigr)/\lambda_\nu\nonumber\\
& + \delta\Delta_F (t) \left[1- 2\frac{d}{d\Delta_F}\delta\omega\right].
\end{align}
To calculate the derivative of $\delta\omega$ in the adiabatic limit we use Eq.~(\ref{eq:delta_omega}). As seen from this equation, $\delta\omega$ is linear in $\Delta_F$. Since $\zeta^{(0}$ is independent of $\Delta_F$, we obtain
\begin{align}
\label{eq:derivative_omega}
\frac{d}{d\Delta_F}\delta\omega = \left(\frac{\gamma_{12}}{\omega_1}\zeta^{(0)}{}^2+\frac{3\gamma_{11}}{2\omega_1}\right)\frac{\Gamma_2\omega_2F_1}{\gamma^{(+)}_{21}\Gamma_2F_1+\gamma^{(+)}_{12}\Gamma_1F_2}.
\end{align} 

In the adiabatic limit the value of $\phi_-$ does not change. Therefore the coefficient of $\delta\Delta_F (t)$ in Eq.~(\ref{eq:adiabat_phi_-}) is equal to zero. Since the first term in the right-hand side of this equation is $-C\delta\Delta_F (t)$, the condition $\dot\phi_-=0$ provides a compact form of the parameter $C$ in terms of $d\delta\omega/d\Delta_F$,
\begin{align}
\label{eq:C_short}
C= 1-2\frac{d\delta\omega}{d\Delta_F}.
\end{align}
An important consequence is that $C$ is independent of the value of $\Delta_F$ itself. Moreover, given that $F_1\propto F_2$, we see that $C$ is independent of the drive amplitude as well.

\section{DEPENDENCE OF THE SETTLING TIME ON PUMP CURRENT}
\label{section:E}

\begin{figure}
\includegraphics[scale=0.65]{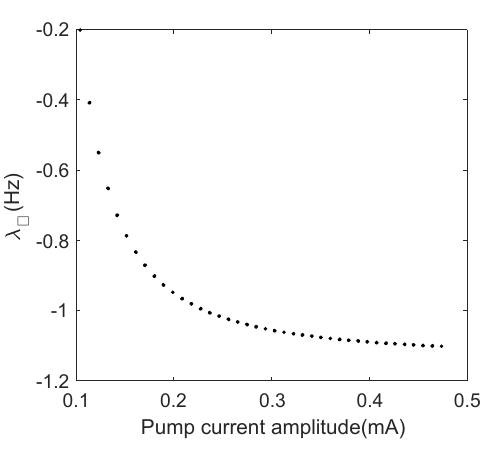}

\caption{Dependence of eigenvalue $\lambda_3$ of matrix $\mathbf{K}$ in the equation of motion linearized about a stable self-sustained vibrations state on the amplitude of the pump current. The pump current frequency detune is 0 Hz.}

\label{Fig.7} 
\end{figure}

The time required for the two modes to settle in response to a step change in $\theta_F$ is determined by relaxation time $|\lambda_3^{-1}|$. In terms of stabilizing the frequencies of self-sustained vibrations, it is desirable to have large $|\lambda_3|$ so that the duration of each cycle in the stabilization algorithm can be reduced and $\theta_F$ can be updated more frequently to compensate for phase diffusion. Figure \ref{Fig.7} plots the dependence of $\lambda_3$ on the amplitude of the pump current when the pump detune frequency is fixed at 0 Hz. Initially, $\lambda_3$ becomes more negative rapidly as the pump current amplitude increases and then levels off. For the data reported in the main text, the pump current amplitude is chosen to be 189 $\mu$A. Further increase in the pump current amplitude only marginally shorten the relaxation time, while significantly increase the risk of damaging the thin gold layer on the suspended beams. 

%%%%%%%%%%%%%%%%%%%%%%%
%%%%%%%%%%%%%%%%%%%%%%%

 %%%%%%%%%%%%%%%%%%%%%%%%%%%%%%%%%%%%%%%%%%%

%\bibliography{re_freq_stab,Zogero23} 


\end{document}